\documentclass[
aps,%
12pt,%
final,%
notitlepage,%
oneside,%
onecolumn,%
nobibnotes,%
nofootinbib,%
superscriptaddress,%
noshowpacs,%
centertags]%
{revtex4}

\begin{document}
\selectlanguage{english}

\title{Search for Fast Radio Bursts in the Direction of the Galaxies M31 and M33}

\author{\firstname{V.~A.}~\surname{Fedorova}}Fedorova V.A.
\email{fedorova@prao.ru}
\affiliation{Pushchino Radio Astronomy Observatory, Astro Space Center, Lebedev Physical Institute, Pushchino, Moscow region, Russia}
\author{\firstname{€.~….}~\surname{Rodin}}Rodin €.…. 
\email{rodin@prao.ru}
\affiliation{Pushchino Radio Astronomy Observatory, Astro Space Center, Lebedev Physical Institute, Pushchino, Moscow region, Russia}


\begin{abstract}

{\bf Abstract} -- The results of a search for individual fast radio bursts with the Large Phased Array of the Lebedev Physical Institute at 111 MHz during July 2012 through August 2018 are presented. The signals were distinguished by convolving the data with a template with a fixed form, followed by convolution with test dispersion measures. Areas of sky containing the galaxies M31 and M33 were chosen for the search. Three radio bursts were detected in the vicinity of M33, five in the vicinity of M31, and one in a region offset from the center of M31 by an hour in right ascension. The dispersion measures of the detected bursts range from 203 to 1262 $ pc \cdot cm^{-3}$.

{\bf DOI:} 10.1134/S1063772919110039
\end{abstract}

\maketitle

\section{INTRODUCTION}

Fast radio bursts (FRBs) are individual pulses with widths from 0.08 to 26 ms. They are likely extragalactic in origin, as is indicated, for example, by the fact that their dispersion measures usually range from 109 to $\sim$ 2600 $pc/cm^3$. Furthermore, there is no
concentration of FRBs toward the Galactic plane in their distribution on the sky (as there is for pulsars) and there are no other distinguished directions in this distribution. According to data from the FRB catalog \cite{cat}, about 100 FRBs are known. One of these,
FRB 121102, has repeated a number of times from 2012 to the present time, at irregular time intervals. Thanks to these repeating bursts, it was possible to identify FRB 121102 with the vicinity of an irregular dwarf galaxy with active star formation \cite{KOKUBO}. This is the only FRB for which a host galaxy has been identified.

A second repeating burst was registered during a survey at 400 MHz using CHIME. Six repeating bursts with dispersion measures $\sim$ 189 $pc/cm^3$ and burst widths from 2.6 to 63 ms were registered from FRB 180814.J0422+73 during August -- October 2018. Twelve new FRBs with various dispersion measures from 169.134 to 1006.84 $pc/cm^3$ were also
detected in this survey \cite{Amiri}.

Extragalactic FRBs have been known since 2007,
when the first such burst was found by chance in
archival data \cite{lorimer}. The first targeted attempts to detect extragalactic burst-like signals were made by Linscott
and Erkes \cite{Linscott} in 1980. Subsequently, McCulloch
et al.\cite{McCulloch}  carried out systematic observations of
the Large Magellanic Cloud in 1980--1981 with the
aim of searching for radio pulsars. Later, attempts
were made to detect giant pulses from pulsars in
various external galaxies, including M33, NGC253,
NGC300, and NGC7793 \cite{McLaughlin}. In 2012, Rubio-Herrera
et al. \cite{Rubio?Herrera} carried out a search for burst signals in the
galaxy M31. This led to the first detection of several
bursts with dispersion measures corresponding to an
extragalactic origin in the direction of M31.

The choice of relatively nearby galaxies for FRB
searches was based on the idea that the objects giving
rise to such bursts should be much more numerous
in densely populated regions of the Universe, such
as spiral galaxies. Two spiral galaxies in the Local
Group, M31 and M33, fall in the field of view of the
Large Phased Array (LPA) of the Pushchino Radio
Astronomy Observatory. Accordingly, we selected
these galaxies as targets for our FRB search.

Since there are a large number of theoretical models
aiming to describe the origin of these mysterious
events, we did not make any a priori assumptions
about the nature of FRBs. We considered a purely
observational problem -- the detection of individual
bursts displaying dispersion delays in frequency in the
directions toward M31 and M33.

The observations were conducted at 111 MHz.
This means that, for potentially detected radio bursts
with dispersion measures of 10$^2$ -- 10$^3$ $pc/cm^3$, the
pulse broadening in an individual channel with width   415 kHz should comprise a few to tens of seconds;
i.e., the bursts being searched for should be very
distorted by the dispersion delay and scattering. It is
also obvious that such broadening will lead to a drop
in the detection sensitivity by a factor of 10$^3$ -- 10$^4$ for
bursts with widths of several millseconds. In spite
of this, astrophysical events with energies $>$ 10$^{46}$ J
occurring in a nearby galaxy should be detectable.
Table 1 presents the parameters of nine new bursts
we have detected in this study. The remaining sections
of the paper describe the technical characteristics
of the LPA, the observing and reduction method
applied to archival data obtained during 2012 -- 2018,
and our results.
	
\section{APPARATUS}

The LPA of the Pushchino Radio Astronomy
Observatory (PRAO) is one of the most sensitive
radio telescopes operating at meter wavelengths. The
operational frequency range of the LPA is 111$\thickspace$$\pm$$\thickspace$1.25 $\thickspace$MHz. The fluctuational sensitivity is 140 mJy
for a time resolution of 0.1 s in a receiver bandwidth
of 2.5 MHz \cite{oreshko}. The signals are registered using a
multichannel digital receiver that enables recording
in two regimes. In the first, the signal is written with
relatively low frequency resolution using six frequency
channels with bandwidths of 415 kHz each. In this
mode, the time interval between readouts is 100 ms.
This regime is used for continuous monitoring of
scintillating sources. The second recording regime
is carried out using 32 frequency channels with
bandwidths of 78 kHz each, with a time resolution
of 12.5 ms. Independent of the recording regime, the
signal is reduced digitally using a 512-readout FFT processor. In our studies of the galaxies M31 and
M33, we used data with a time resolution of 100 ms.

In connection with the specific features of the
directional beam of the LPA, the effective area of
the anntenna has its maximum value at the zenith
(47 000 m$^2$) and decreases toward the horizon proportional
to cos $z$, where $z$ is the zenith distance.
The field of view of the radio telescope is $\sim$ 50 square
degreees. This enables daily monitoring of a large
number of sources. The system noise temperature
depends on the sky background and varies from 550
to 3500 K.

A distinguishing characteristic of the LPA meridian
radio telescope is its directional beam (DB), which
includes a beam with an adjustable declination offset
(DB-1) and a stationary beam (DB-3). DB-1 is used
for studies of pulsars; the LPA has successfully been
used to find pulsars by monitoring an appreciable
fraction of the sky around the clock \cite{rodin}. Observations
with DB-3 can be carried out continuously.
Since FRBs are sporadic events, our monitoring was
conducted using DB-3, which has 96 sub-beams and
was designed for studies of interplanetary scintillations
of a large number of compact radio sources. The
sub-beams cover the sky in declination from $-9^{\circ}$ to $42^{\circ}$. The half-width of the main DB-3 beam varies
from $24^\prime$ to $48^\prime$,depending on the declination of the
observed source.

M33 and M31 are extended sources, with angular
sizes of $73^\prime$ å $45^\prime$ (Œ33) and $3.2^{\circ}$ å $1^{\circ}$ (Œ31). Such
extended sources in the LPA field of view are covered
by several sub-beams of the main DB. Since M31
has declination $\delta_{J2000} = 41.27^{\circ}$, observations of this
source were carried out using sub-beams 1--6 of  DB--3. For M33, $\delta_{J2000} = 30.66^{\circ}$, corresponding to
observations using sub-beams 27 and 28 of DB-3.
Figure 1 presents images of the studied galaxies in
the optical together with the declinations of the DB--3
sub-beams.

The interference environment around the LPA is
monitored regularly. Based on many years of measurements,
several types of interference have been
distinguished:
\begin{enumerate}
\item Atmospheric (lightning discharges, perturbatios
of the ionosphere during solar flares);
\item Industrial (spark discharges from electrical
instruments, passing automobiles, electric welding,
and others);
\item Interference from radio equipment (television,
VHF radio stations, radar, and others).
\end{enumerate}

From the point of view of their effect on the LPA,
all of these sources of interference have a common
property: although they may be spatially localized,
they influence the LPA receiver tract as a whole. This
means that the interference signal arises simultaneously
in several or even all sub-beams at the antenna
output. There is no dependence of the arrival
times of the interference signals on frequency (they
exhibit zero dispersion measure) and interference is
extended over time. This is true for both industrial
interference and interference associated with solar
ionospheric perturbations, which are registered in all
sub-beams, and even in the antenna sidelobes. Thus,
in contrast to terrestrial interference, a cosmic pulsed
signal arrives from a specific direction on the sky and
displays a frequency dependence for the arrival time.

Interference displaying frequency dependence of
the arrival times, imitating a cosmic dispersion signal,
is sometimes registered. However, in this case, the
signal power exceeds the power of a cosmic signal very substantially, and the interference signal is registered
in several sub-beams, making it easily distinguishable.

Overall, the interference environment worsens
with the onset of Summer and the thunderstorm
season, leading to an increase in the number of
spurious signals in the data.

We considered pulses satisfying the following conditions
to be candidate FRBs:
\begin{enumerate}
\item Visible in a single sub-beam,
\item Displaying a frequency dependence for the arrival
times,
\item Registered in all six channels.
\end{enumerate}
Pulses not satisfying these conditions were taken
to be interference and were not considered to be candidate
FRBs.

\section{DATA REDUCTION}

The data were reduced as follows. We used mathematical
modeling to obtain the pulse-like signals
expected for FRBs, taking into account their propagation
in the interstellar medium and detection at low
frequencies. Our modeling of the expected signals
took into account the fact that the received burst
differs from the emitted burst. The pulse shape is
distorted by the influence of the inhomogeneous interstellar
medium through which it propagates, and
the resulting scattering $t_s$ should be correlated with
its dispersion measure $DM$. This correlation is described
by Kuz'min et al.  \cite{kuz} and is given by $t_s=0.06(\frac{DM}{100})^{2.2}$ at 110 MHz. In addition, the pulse
undergoes a dispersion delay in frequency. The registration
of the signal in a finite frequency band also
leads to broadening, which is described by convolving
the received signal with a $\Pi$-like function:

\begin{equation}
\Pi (t) = \sigma(t - \tau_{i-1})\sigma(\tau_i - t)
\end{equation}
where $\sigma(t)$ -- is a single step function and $\tau_i$ -- is the
arrival time at the boundary of frequency channel $i$. The quantity $\Delta\tau=\tau_i -\tau_{i-1}, (i = 1,2,3,\dots, 6)$ is the
broadening of the pulse within the band. To distingish
impulsive dispersion signals, we convolved a pulse
containing noise with a template applying dispersion
compensation, making it possible to obtain the maximum
signal-to-noise ratio (SNR). The process used
to conduct the modeling and obtain the template is
described in detail in \cite{Fedorova}.

The method used to search for FRBs was as follows.
We first analyzed the daily recordings in the
six frequency channels with a time resolution of 0.1 s
in sub-beams coinciding with the directions toward
M31 and M33. We then applied corrections taking
into account the deviation of the LPA sub-beams
from the plane of the celestial meridian and precession.
For M33, in sub-beams 27 and 28, we selected
the five-minute section from an hourly recording corresponding
to the right ascension of M33 ($\alpha_{J2000} = 01^h 34^m$). In the case of M31, we chose a half-hour
section of a recording in the first six sub-beams, with
their center corresponding to the right ascension of
the center of M31 $\alpha_{J2000} = 00^h 43^m$, Further, we convolved
the recordings in each sub-beam with a template
obtained through our mathematical modeling,
after which we convolved the resulting signal with test
dispersion measures ranging from 0 to 3000 $pc/ám^3$ in steps of 50 \cite{Fedorova}. 

Since FRBs manifest as weak individual pulses
at 111 MHz, due to their appreciable scattering and
broadening, it is not possible to directly detect these
signals without applying additional methods such as
we have used. Our approach implementing convolution
with a template with an appropriate form makes it
possible to amplify the burst signal against the noise
background. As an example of the operation of the
method, Fig. 2 presents the same section of recording
before and after convolution with the template.  

The subsequent processing of the data amounted
to a visual analysis of the results obtained in the last
stage of the reduction. When a burst signal was
detected, the dispersion measure, peak flux density,
and SNR were determined separately in each case.
The results are presented in Table 1.

\section{RESULTS}

Our visual analysis of the data led to the detection
of nine FRBs with dispersion measures from 203 to
1262 $pc/ám^{3}$ (see Figs. 3 -- 11).   One of these may
have repeated, as is suggested by two bursts with coordinates
coincident within the half-width of the LPA
direction beam and displaying dispersion measures
coincident to within $\pm 4$ $pc/ám^{3}$. These tentatively
repeating bursts were registered several days apart
(November 25, 2015 and November 28, 2015).

The SNR values for the registered bursts after
applying a matched filter were all $\lesssim 10$, indicating that
the peak flux densities for the registered bursts given
in Table 1 are at the limit of the antenna sensitivity.
Note also that we distinguish only the upper part of
the burst with our method, since the lower, exponentially
decaying, part is lost in the noise.

The burst registered on March 21, 2018 (Fig. 10)
merits special attention. The main signal is clearly
visible in the dynamical spectrum, however another
pulse can be noted to the right of the main pulse after
$\sim$300 readouts. The second pulse is clearly visible
in the right-hand part of the total recording for six
frequency channels. This event is of special interest
and requires further more detailed study.

Table 1 also presents the right ascension corresponding
to the sixth frequency channel ($f = 111.5$ MHz) and the estimated SNR.

The dispersion measures of Galactic pulsars range
from $\sim\thickspace3$ to $\sim 1800$ $pc/ám^{3}$ for pulsars in the direction
of the Galactic center, which fully covers the range of
dispersion meaures for our detected FRBs. Some of the detected pulses with high $DM$ values are far from
the optically bright parts of M31 and M33, where the
concentration of matter is appreciably lower than in
the central regions. This suggests that, since a sufficient amount of matter to explain the observed $DM$
values is not accumulated in the regions where the
bursts are located, not all the bursts are associated
with the two galaxies studied, and more likely coincide
with the directions of these galaxies by chance.

The presence of noise in the recordings on some
days could suggest that the detected bursts could
form by chance due to the superposition of intensity
variations at "needed" places in the recordings in the
six frequency channels. Therefore, we decided to
subject the results obtained to a detailed statistical
analysis. This analysis was carried out in two ways:
we calculated the probability of a chance alignment
of the pulses to form the dynamical spectra and the
probability that the burst arose by chance, as functions
of the SNR.

Recall that we analyzed daily recordings with
durations of $T_{rec}$$\thickspace$=$\thickspace$300$\thickspace$s. Each recording was
passed through a matched filter with characteristic
width $t_s$$\thickspace$=$\thickspace$1$\thickspace s$s, leading to $m=T_{rec}/t_s$. independent
readouts in the recording. Figure 12 schematically
shows the dynamical spectrum with all quantities
used. We introduced the quantity $k=t_{pulse}/t_s$. The widths of the detected pulses $t_{pulse}$ are several
seconds, yielding $k\sim$$\thickspace$1--5. We also introduced
the quantity $p=T_{rec}/t_{pulse}$. The probability that
readings in one channel line up, forming a pulse,
is $P_m=1/(m-1)(m-2)...(m-k)$. The probability
that individual bursts in channels form the dynamical
spectrum is $P_n=1/p^{(n-1)}$, where $n=6$ is the number
of frequency channels. The --1 in the power-law
index in the denominator for the probability $P_n$ takes
into account the fact that a pulse can appear in any
place in a recording, and that this does not affect the
fact of a detection. Thus, the overall probability that the readouts align to form the observed dynamical
spectrum by chance is $P_{tot}=P_m \cdot P_n\sim10^{-12}$--$10^{-19}$.

The probability that a readout exceeds a threshold
SNR was calculated as follows. We subtracted mean
values smoothed over 150 readouts obtained using
a median filter from the recordings in each channel.
We further constructed the empirical distribution of
the readouts. This distribution was best described by
a Laplace (double exponential) distribution. We determined
the parameters of the distribution (its mean
and dispersion), and used these parameters to calculate
the probability of exceeding a specified SNR. This
probability proved to be $P_r=3\cdot10^{-2}\div8\cdot10^{-3}$ for SNR $\sim7 \div10$.

Thus, the detection of FRBs reduces to simultaneously
satisfying two conditions: the readings exceeding
a threshold SNR and alignment of these readings
to form the dynamical spectrum. The probability that
both of these conditions are satisfied simultaneously
by chance is $P = P_r\cdot P_{tot} \sim 10^{-14}\div10^{-23}$; in our view,
this is too low to occur by chance in 6930 hrs of
analyzed data in  $\sim82 \thickspace 000$ individual scans.

\section{DISCUSSION}

Our primary aim was to search for and estimate
the parameters of FRBs in the directions of the galaxies
M31 and M33 in archival LPA data. The estimated
dispersion measures for the detected bursts presented
in Table 1 and the absence of a visible concentration
of the bursts toward the centers of the two galaxies
studied lead to the natural suggestion that the regions
in which some bursts were generated are located far
beyond these galaxies. Thus, we conclude that M31
and M33 are not the host galaxies for all the registered
burst signals

We registered similar bursts in other parts of the
sky earlier  \cite{Rodin}. The characteristics of our newly
detected signals are similar to those of the pulses reported in \cite{Fedorova}, and also to the cataloged parameters
of FRBs.

We separately touch on the question of the observed
number of radio bursts. A region of sky covering $\sim 30$ square degrees observed over six years was
analyzed. Over this time, nine bursts were detected.
Recalculating this number to the entire area of the
sky and supposing that the detected bursts are not
associated with the galaxies M31 and M33, we obtain
a mean burst detection rate for the LPA at 110 MHz
of $\sim 2000$ bursts/year.

At present, one problem with studies of these
events is estimates of their spectral indices. Estimates
of this quantity for FRB 121102 in various
articles range from 1 to 10 \cite{Spitler}. In our view, this
large scatter in the deduced spectral indices can be
explained mainly by the fact that these measurements
were carried out in a receiver bandwidth in which
the instantaneous frequency distribution of the peak
flux density is determined by scintillation as the pulse propagates in the interstellar medium, not physics
associated with the radiation itself. After detecting a
large number of new bursts from FRB 121102 during
2018, Macquart et al. \cite{Macquart} derived the spectral index $\alpha$ = -- 1.6. In \cite{aTel}, we reported the spectral index
for FRB121102 $\alpha\sim\thickspace-0.6 \pm 0.4$. This result was
based on an analysis of measured flux densities for
the repeating radio burst FRB121102 and on our
suggestion \cite{Fedorova} that we registered and measured a
repeat pulse from this object at 111 MHz.

Egorov and Postnov \cite{egorov} proposed a mechanism
for the generation of radio bursts based on the interaction
of a plasma flow streaming out from the
magnetic poles of a neutron star and a shock that
arises after the SN explosion and passes by the pulsar.
At present, this is not generally considered to be the
main mechanism at work, since it does not explain
accompanying behavior in other ranges or repeating
radio bursts. Nevertheless, the idea of the "ignition"
of matter by a cone of radiation from a pulsar is interesting. If we consider a hypothetical binary
system with a pulsar in which the stellar companion
displays activity in the form of matter ejections, this
mechanism will give rise to radio bursts at irregular
time intervals.

We also wish to mention an idea concerning the
high dispersion measures of radio bursts. Depending
on the relative geometry for the shock, neutron star,
and observer, a situation can arise when a pulse must
pass through a substantial thickness of ionized matter.
In turn, this leads to high values for the observed
$DM$. In this case, the inferred distances at which
radio bursts are believed to arise and the deduced
burst energies must both be lowered.

Returning to the question of modeling the events
generating FRBs, we propose the following.

First, of the entire list of FRBs, two stand out:
FRB121102 and FRB180814.J0422+73. Their
pulses were detected more than once, in contrast. to the others. This already suggests that repeating
and non-repeating FRBs are generated by different
mechanisms.

Second, a visual analysis of the registered FRBs
shows that they have very different appearances,
in particular, different widths, which cannot be
epxlained purely by broadening in the receiver bandwidth
and are associated with internal properties of
the pulses.

Third, the high dispersion measures $DM$ and
estimated energies of all these events may indicate
that we are dealing with a fundamentally new phenomenon
that is not yet known or has been predicted
only theoretically. Possibilities include, for example,
the model proposed in \cite{Tkachev}, in which the Primakov
process is used to explain the transformation of
axions---dark matter particles ---into photons in a
magnetic field. Note that the disruption of clouds of
such matter by an object with a powerful magnetic field could generate pulses with durations of several
seconds \cite{Pshirkov}. All the detected bursts have characteristics
widths of several seconds. Compensation
for scattering in the medium and broadening in the frequency channels can reduce the burst widths to $\sim$ 0.1 -- 2\thickspaceá.

Thus, in spite of the existence of models that can
explain both individual and repeating FRBs, we find the possibility that these two types of events have
different natures more natural.

\section{CONCLUSION}

Our main results follow
\begin{enumerate}
\item We have detected three FRBs in the direction
of M33 in archival LPA data obtained from
July 2012 through August 2018, two of which
tentatively arose in the same region on the sky
and thus may represent a repeating FRB.
\item In this same period, we detected six FRBs with
various dispersion measures in the direction of
M31.
\end{enumerate}

Thus, in all, we detected nine FRBs with dispersion
measures ranging from 203 to 1262 $pc/ám^{3}$ in archival LPA data for the period from July 2012
though August 2018. All the parameters of these
bursts are presented in Table 1.

\begin{center}
ACKNOWLEDGMENTS
\end{center}

We thank the Director of the Pushchino Radio
Astronomy Observatory R.D. Dagkesamanskii and
the Assistant Director V.V. Oreshko for discussions
of this paper that have led to its improvement.

\begin{center}
FUNDING
\end{center}

V. A. Fedorova was partially funded by the Russian
Foundation for Basic Research (grant 16-29-13074).

$Translated$ $by$ $D.$ $Gabuzda$

\appendix

\selectlanguage{english}

\newpage
\begin{center}
{\bf Table 1.} Parameters of detected bursts
\end{center}

\begin{tabular*}{\linewidth}{|p{2.7cm}|p{2.8cm}|p{3.1cm}|p{1.8cm}|p{2.9cm}|p{2.3cm}|} \hline
Date & Coordinates (J2000), $\alpha, \delta$ & DM, $pc/ám^3$ & SNR & $F_max$, Jy & E, Jy $\cdot$ ms \\ \hline
29.10.2012 & 0012 +42.06 & 732 $\pm$ 5 & 7.3 & 0.34 & 1380   \\ \hline
30.10.2013 & 0025 +39.98 & 203 $\pm$ 4 & 10.1 & 0.24 & 800   \\ \hline
12.02.2014 & 0131 +30.54 & 910 $\pm$ 4 & 9.2 & 0.26 & 945   \\ \hline
16.12.2014 & 0014 +41.64 & 545 $\pm$ 5 & 7.6 & 0.23 & 1200  \\ \hline
25.11.2015 & 0131 +30.98 & 273 $\pm$ 4 & 8.5 & 0.54 & 2450  \\ \hline
28.11.2015 & 0132 +30.98 & 273 $\pm$ 4 & 7.2 & 0.52 & 2360  \\ \hline
06.02.2016 & 0101 +41.63 & 1262 $\pm$ 5 & 7.9 & 0.26 & 1780 \\ \hline
02.12.2016 & 2344 +40.80 & 291 $\pm$ 4 & 7.1 & 0.29 & 1320  \\ \hline
21.03.2018 & 0033 +42.03 & 596 $\pm$ 5 & 8.2 & 0.54 & 2310  \\ \hline
\end{tabular*}

\newpage
\begin{figure}[h!]
\begin{minipage}[h]{0.7\linewidth}
\center{\includegraphics[width=1.1\linewidth]{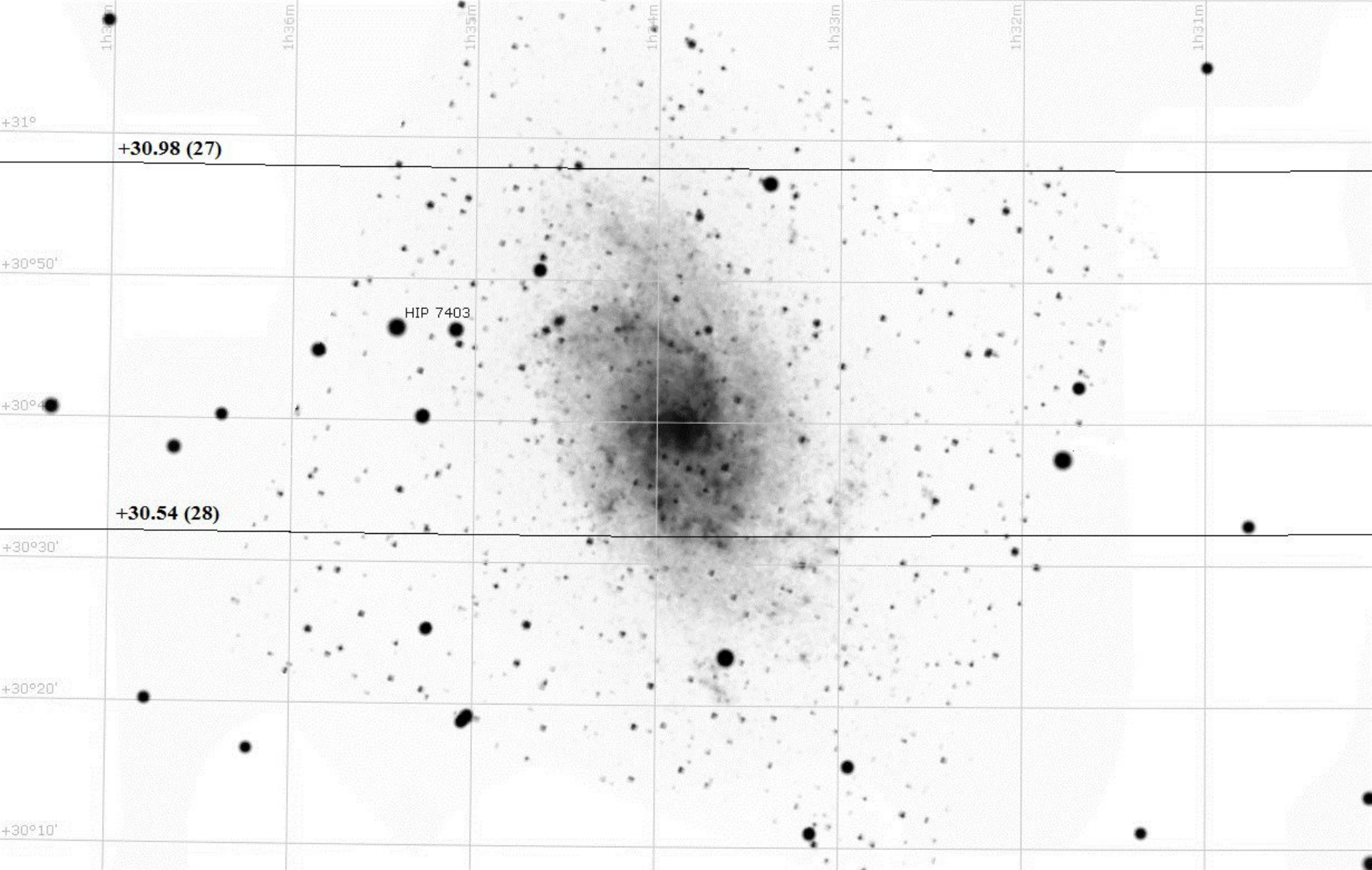}  )}
\end{minipage}
\vfill
\setcaptionmargin{1mm}
\begin{minipage}[h]{0.7\linewidth}
\center{\includegraphics[width=1.1\linewidth]{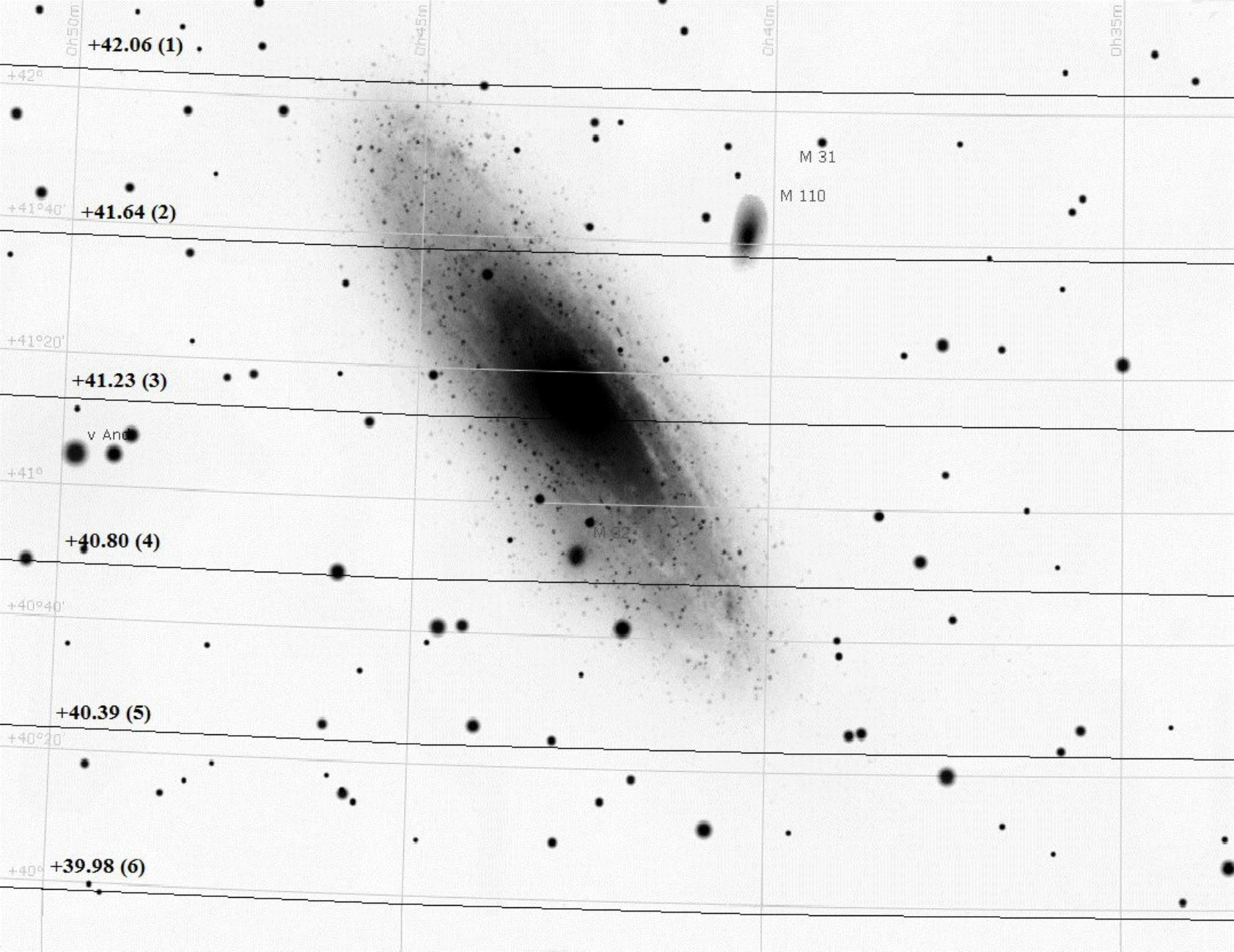} b)}
\end{minipage}
\label{ris:image1}
	\caption{The galaxies M33 (upper) and M31 (lower). The solid black lines show the declinations of the maxima of the LPA
	sub-beams. Each sub-beam is denoted using a coordinate and number corresponding to its location in the DB-3 directional
	beam of the LPA. The coordinates of the sub-beams and galaxies are given for epoch J2000.}
	\label{ris:fig1}
\end{figure}

\newpage
\begin{figure}[h!]
\begin{minipage}[h]{0.7\linewidth}
\center{\includegraphics[width=1\linewidth]{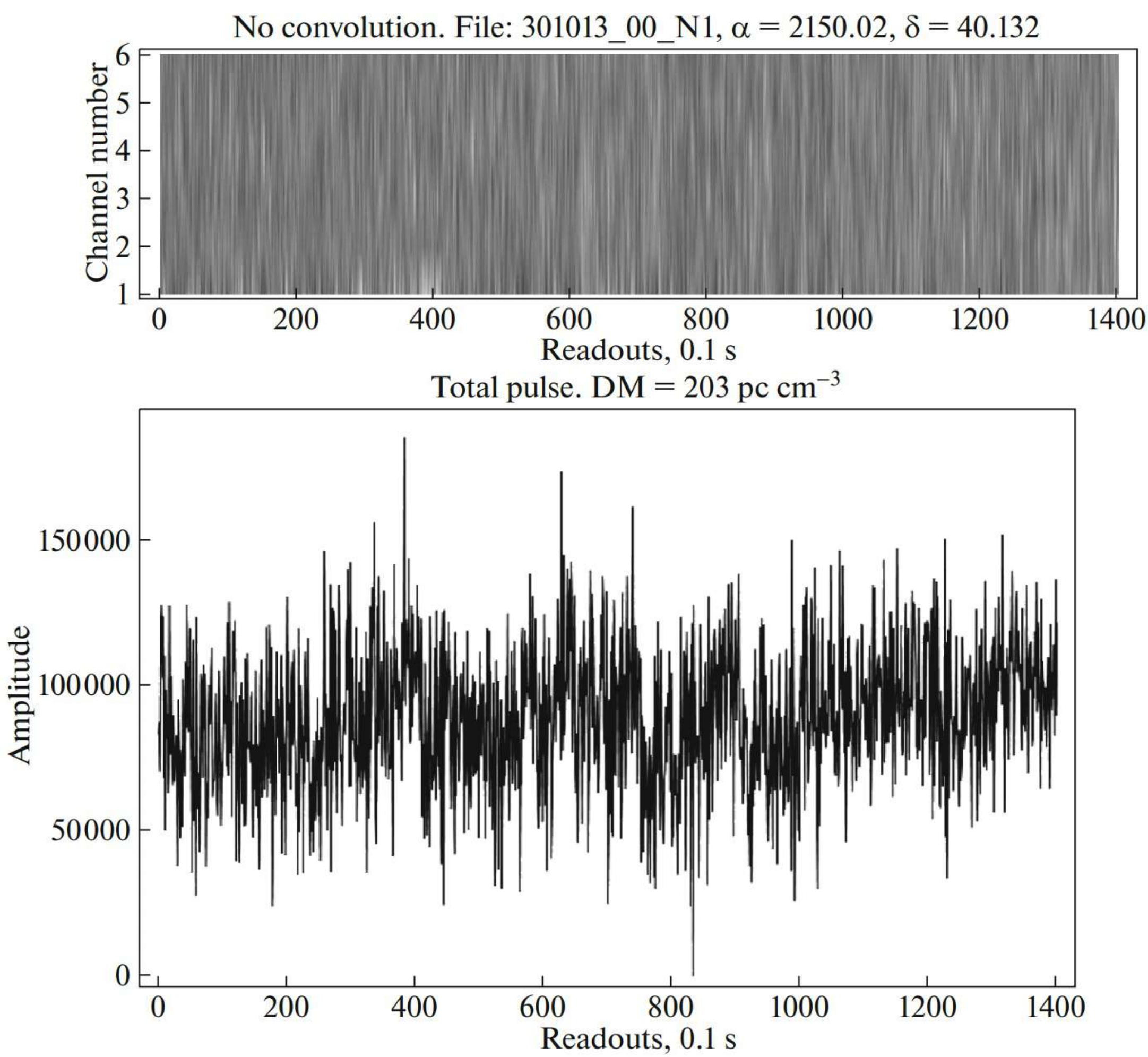} a)}
\end{minipage}
\vfill
\setcaptionmargin{1mm}
\begin{minipage}[h]{0.7\linewidth}
\center{\includegraphics[width=1\linewidth]{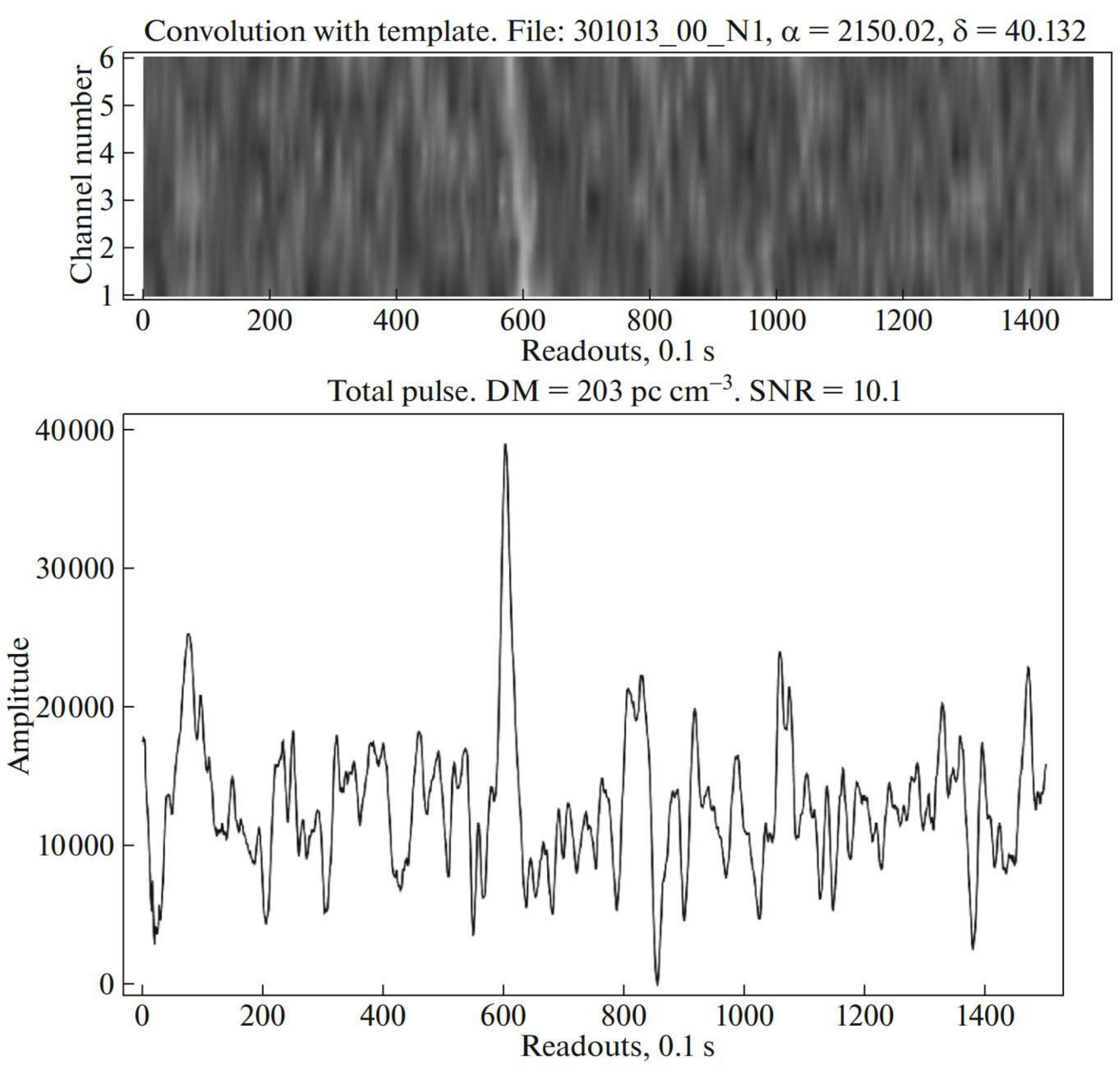} b)}
\end{minipage}
\label{ris:image1}
	\caption{LPA recording with a duration of 150 s corresponding to the FRB detected on October 30, 2013. The same section is
	shown without processing (upper two fragments) and after convolution with the template (lower two fragments).}
	\label{ris:fig2}
\end{figure}

\newpage
\begin{figure}[h!]
	\setcaptionmargin{1.3mm}
	\vbox{\includegraphics[width=0.9\linewidth]{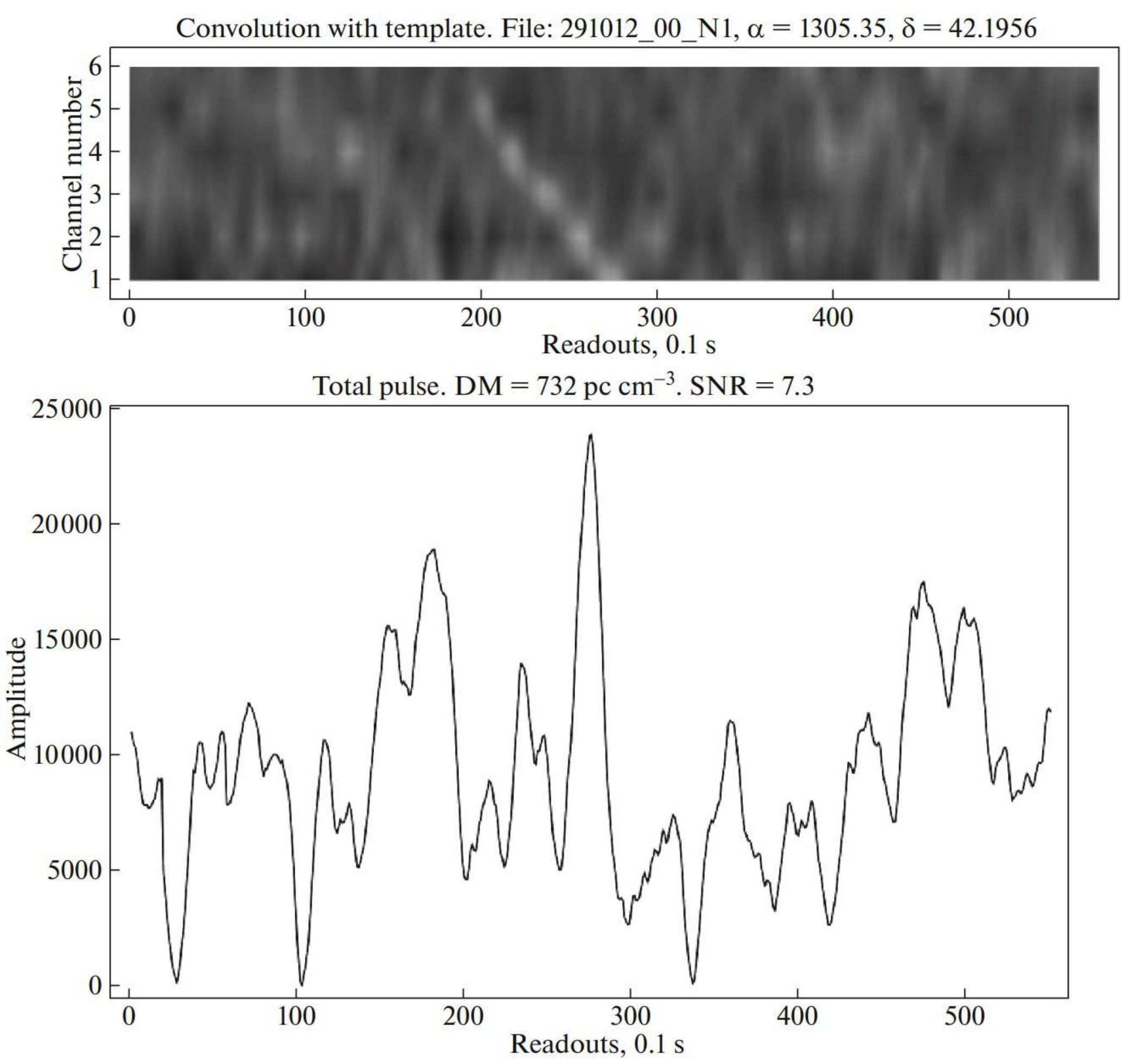}}
	\caption{Dynamical spectrum of the burst with $DM = 732$ $pc/cm^3$ detected on October 29, 2012. The lower image shows the
	total profile of the pulse. The peak flux density is 0.34 Jy. The Galactic coordinates are $l = 115.32^\circ$, $b = -20.24^\circ$.}
	\label{ris:fig1}
\end{figure}

\newpage
\begin{figure}[h!]
	\setcaptionmargin{1.3mm}
	\vbox{\includegraphics[width=0.9\linewidth]{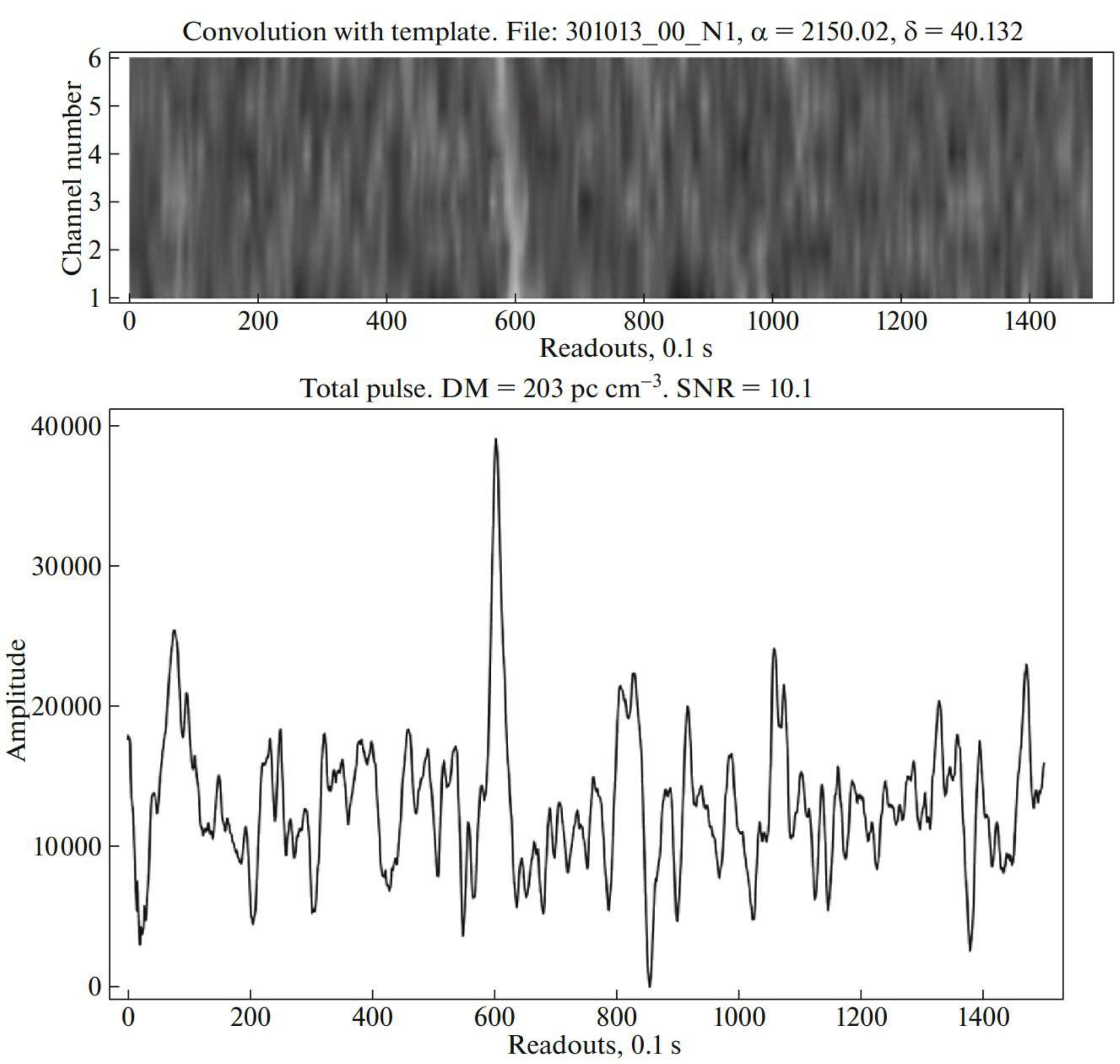}}
	\caption{ Dynamical spectrum of the burst with $DM = 203$ $pc/cm^3$ detected on October 30, 2013. The lower image shows the
	total profile of the pulse. The peak flux density is 0.24 Jy. The Galactic coordinates are $l = 117.53^\circ$, $b =$ -- 22.61$^\circ$.}
	\label{ris:fig1}
\end{figure}

\newpage
\begin{figure}[h!]
	\setcaptionmargin{1.3mm}
	\vbox{\includegraphics[width=0.9\linewidth]{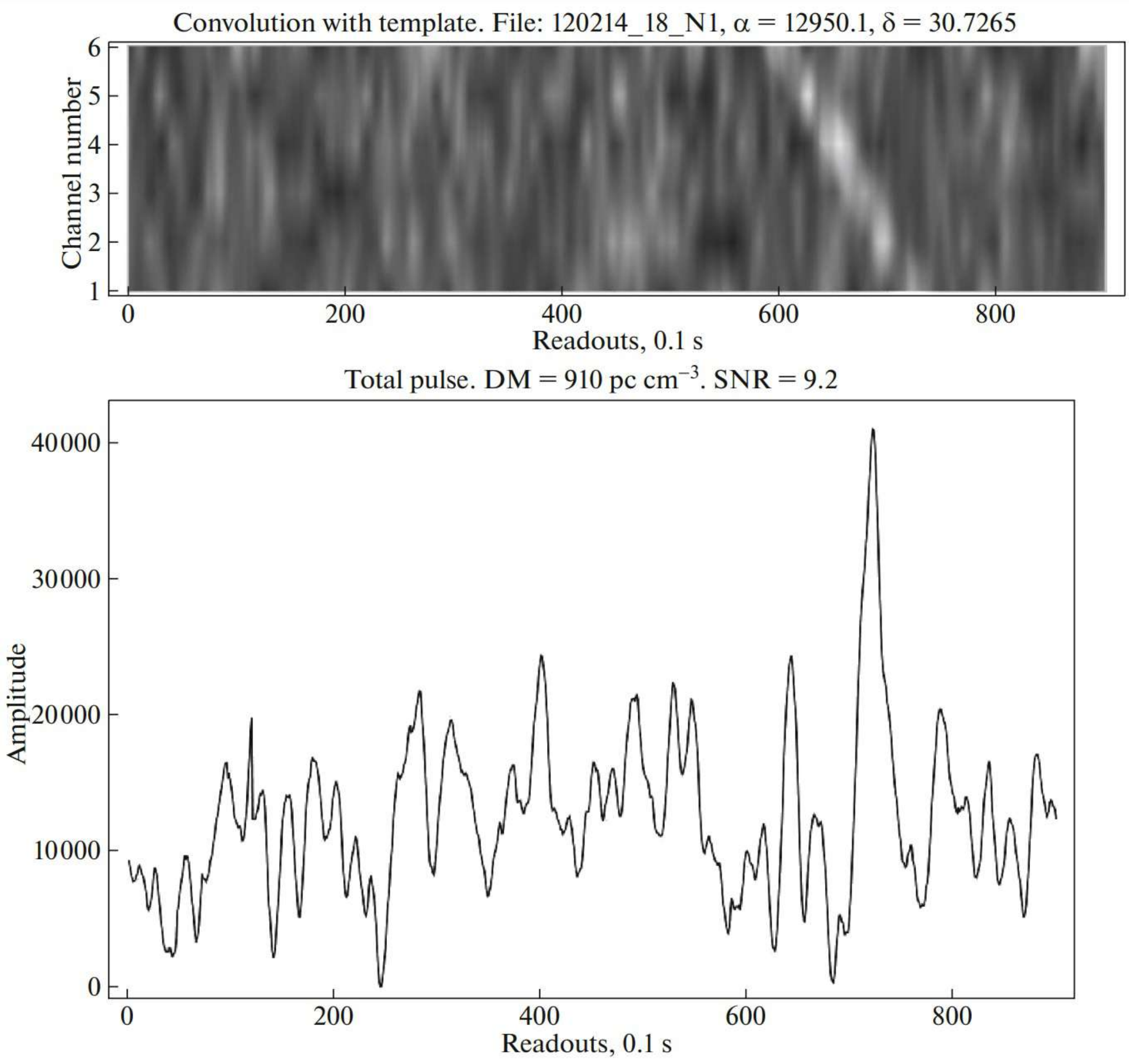}}
	\caption{ Dynamical spectrum of the burst with $DM = 910$ $pc/cm^3$ detected on February 12, 2014. The lower image shows the
	total profile of the pulse. The peak flux density is 0.26 Jy. The Galactic coordinates are $l = 133.10^\circ$, $b =$ -- 31.54$^\circ$.}
	\label{ris:fig1}
\end{figure}

\newpage
\begin{figure}[h!]
	\setcaptionmargin{1.3mm}
	\vbox{\includegraphics[width=0.9\linewidth]{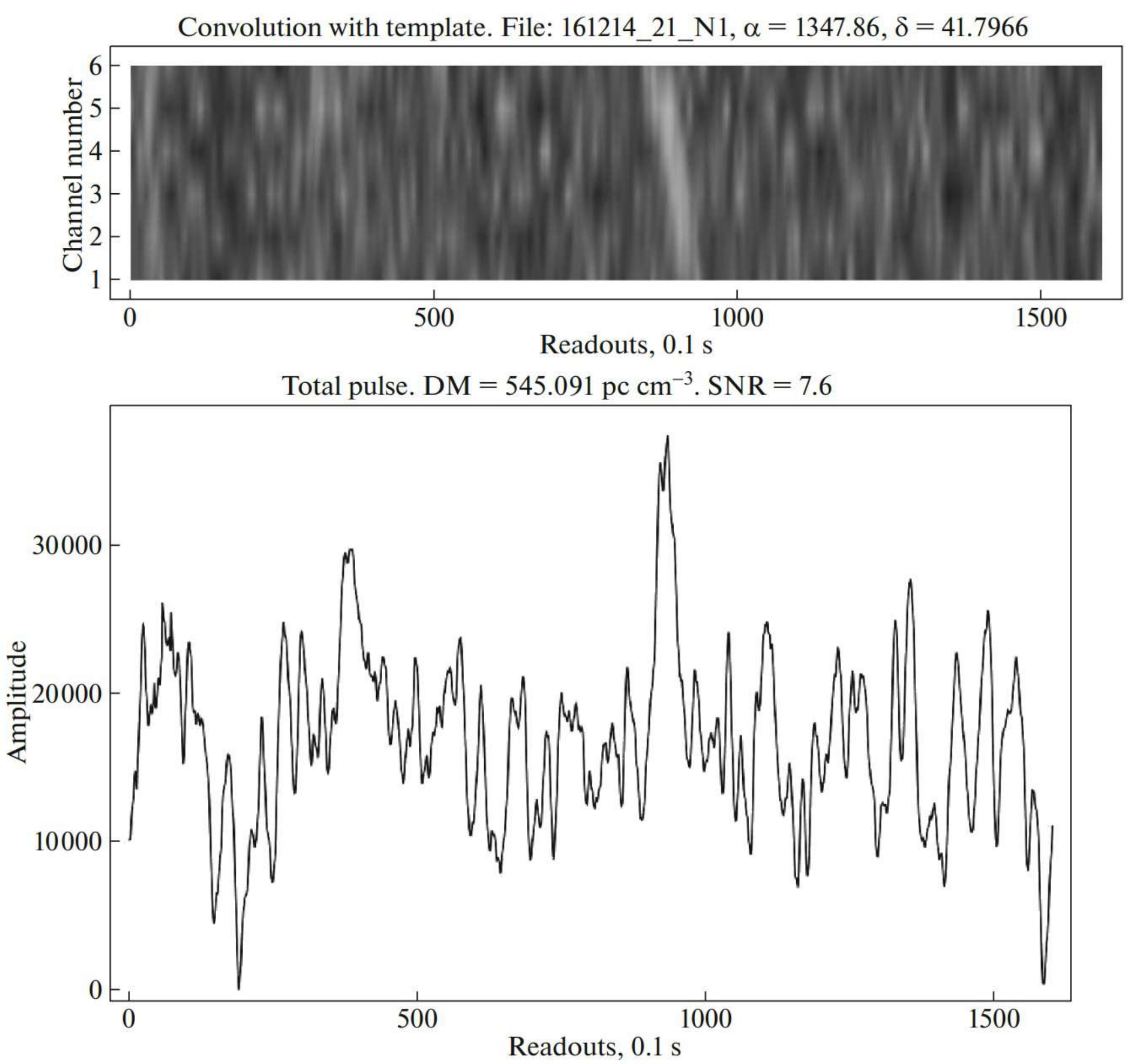}}
	\caption{ Dynamical spectrum of the burst with $DM = 545$ $pc/cm^3$ detected on December 16, 2014. The lower image shows
	the total profile of the pulse. The peak flux density is 0.23 Jy. The Galactic coordinates are $l = 115.55^\circ$, $b =$ -- 20.70$^\circ$.}
	\label{ris:fig1}
\end{figure}

\newpage
\begin{figure}[h!]
	\setcaptionmargin{1.3mm}
	\vbox{\includegraphics[width=0.9\linewidth]{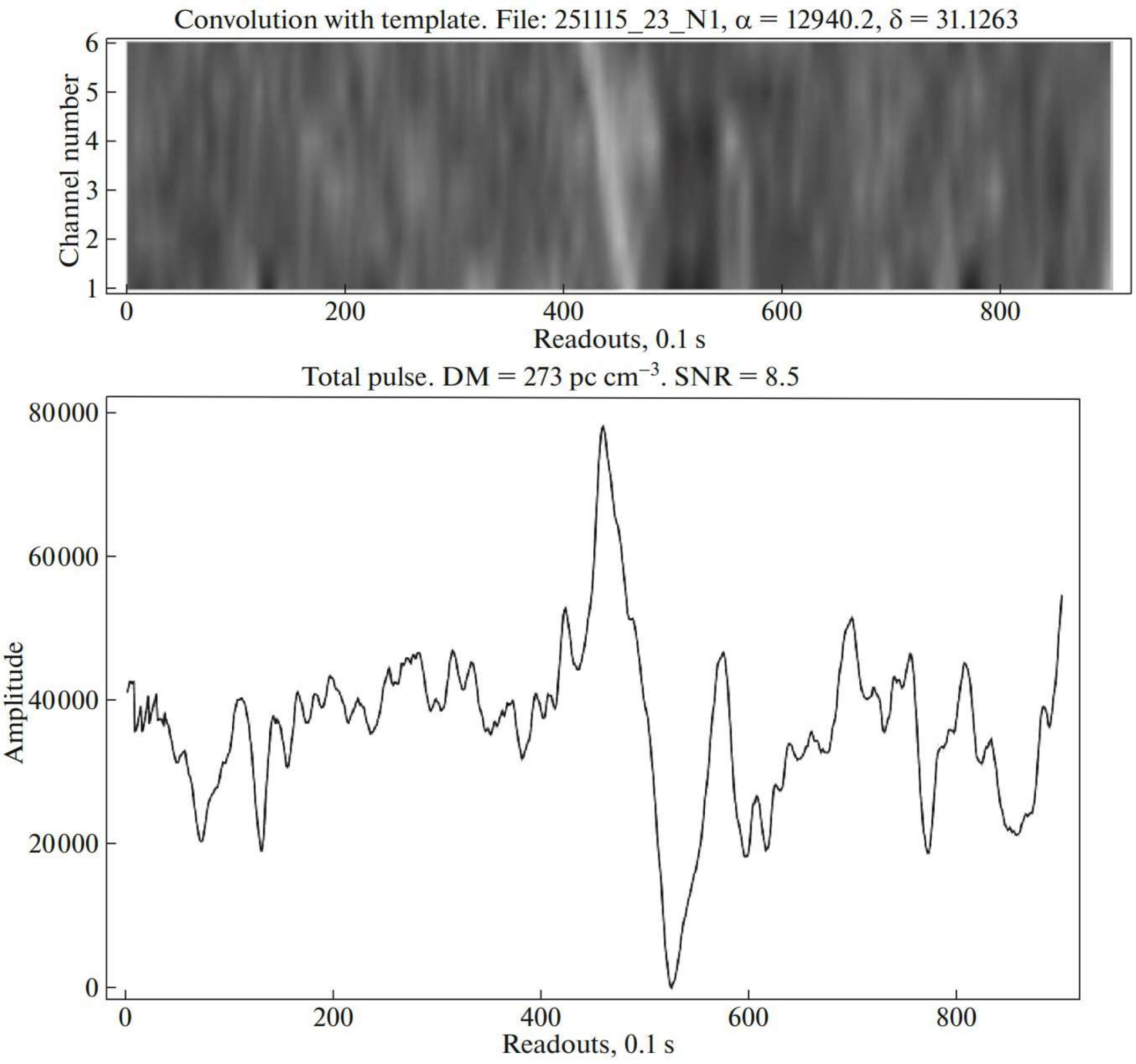}}
	\caption{Dynamical spectrum of the burst with $DM = 273$ $pc/cm^3$ detected on November 25, 2015. The lower image shows
	the total profile of the pulse. The peak flux density is 0.54 Jy. The Galactic coordinates are $l = 133.02^\circ$, $b =$ -- 31.1$^\circ$.}
\end{figure}

\newpage
\begin{figure}[h!]
	\setcaptionmargin{1.3mm}
	\vbox{\includegraphics[width=0.9\linewidth]{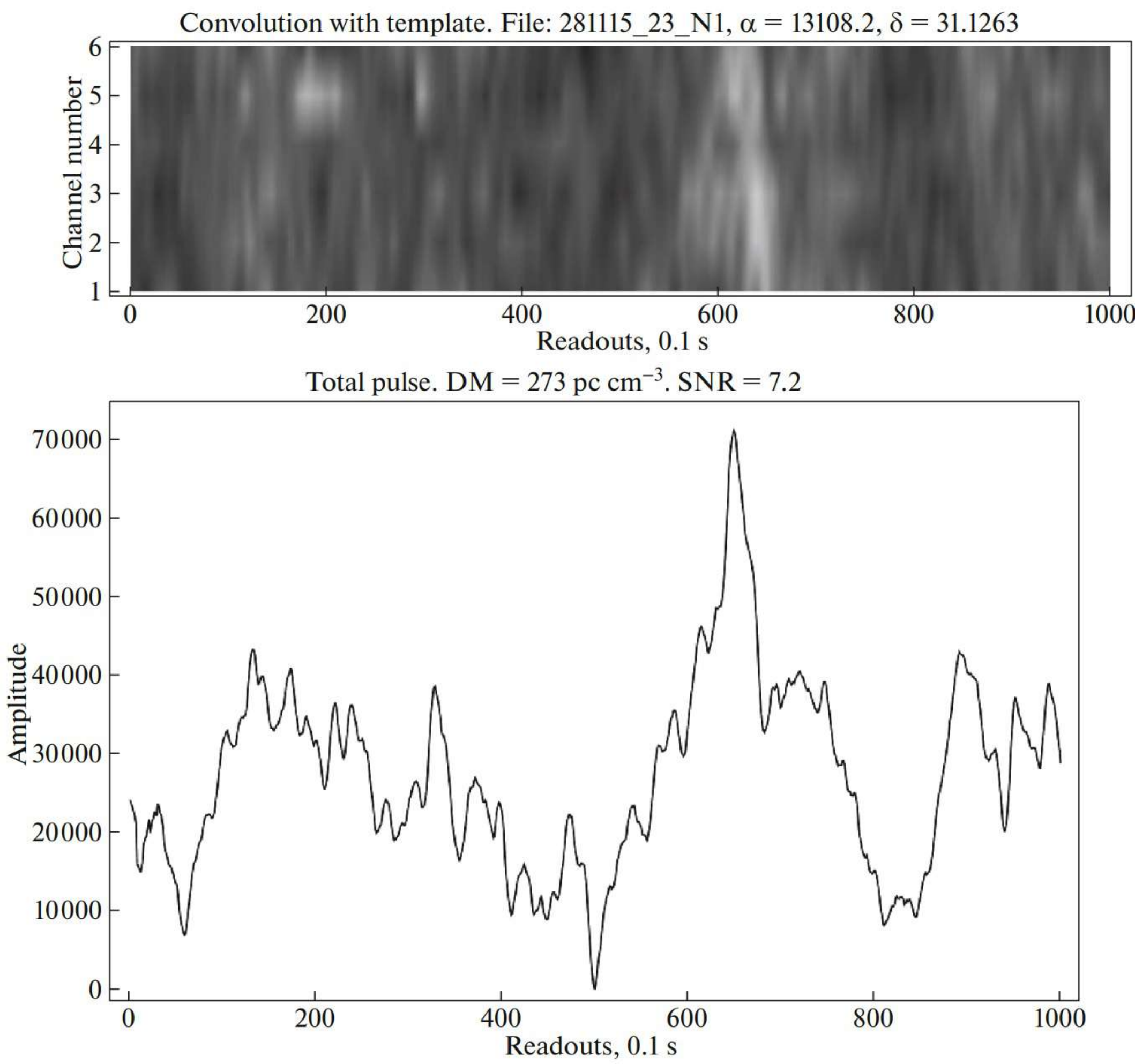}}
	\caption{Dynamical spectrum of the burst with $DM = 273$ $pc/cm^3$ detected on November 28, 2015. The lower image shows
	the total profile of the pulse. The peak flux density is 0.54 Jy. The Galactic coordinates are $l = 133.20^\circ$, $b =$ -- 31.07$^\circ$.}
	\label{ris:fig1}
\end{figure}

\newpage
\begin{figure}[h!]
	\setcaptionmargin{1.3mm}
	\vbox{\includegraphics[width=0.9\linewidth]{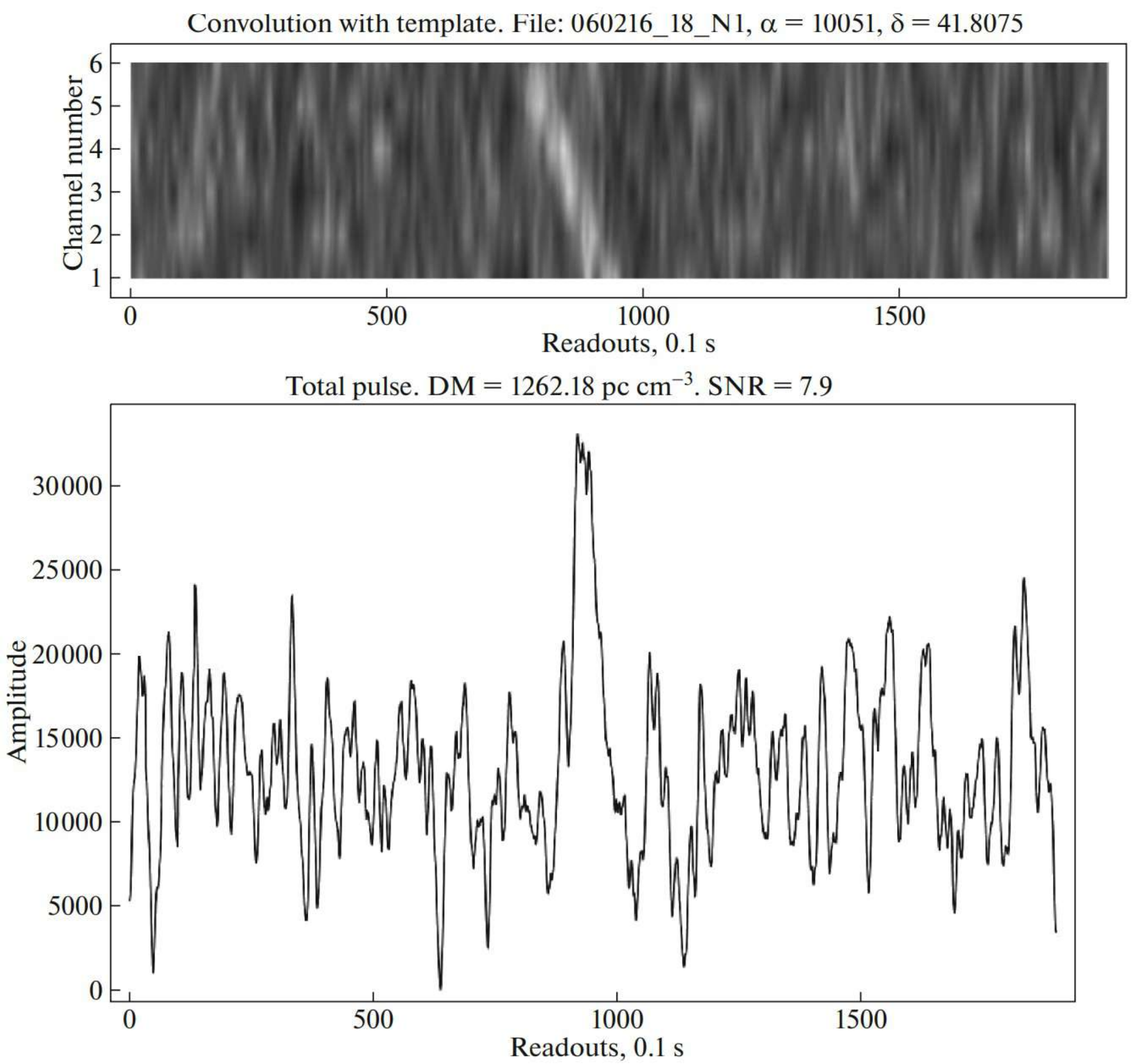}}
	\caption{Dynamical spectrum of the burst with $DM = 1262$ $pc/cm^3$ detected on February 6, 2016. The lower image shows the
	total profile of the pulse. The peak flux density is 0.26 Jy, The Galactic coordinates are $l = 124.90^\circ$, $b =$ -- 21.20$^\circ$.}
	\label{ris:fig1}
\end{figure}

\newpage
\begin{figure}[h!]
	\setcaptionmargin{1.3mm}
	\vbox{\includegraphics[width=0.9\linewidth]{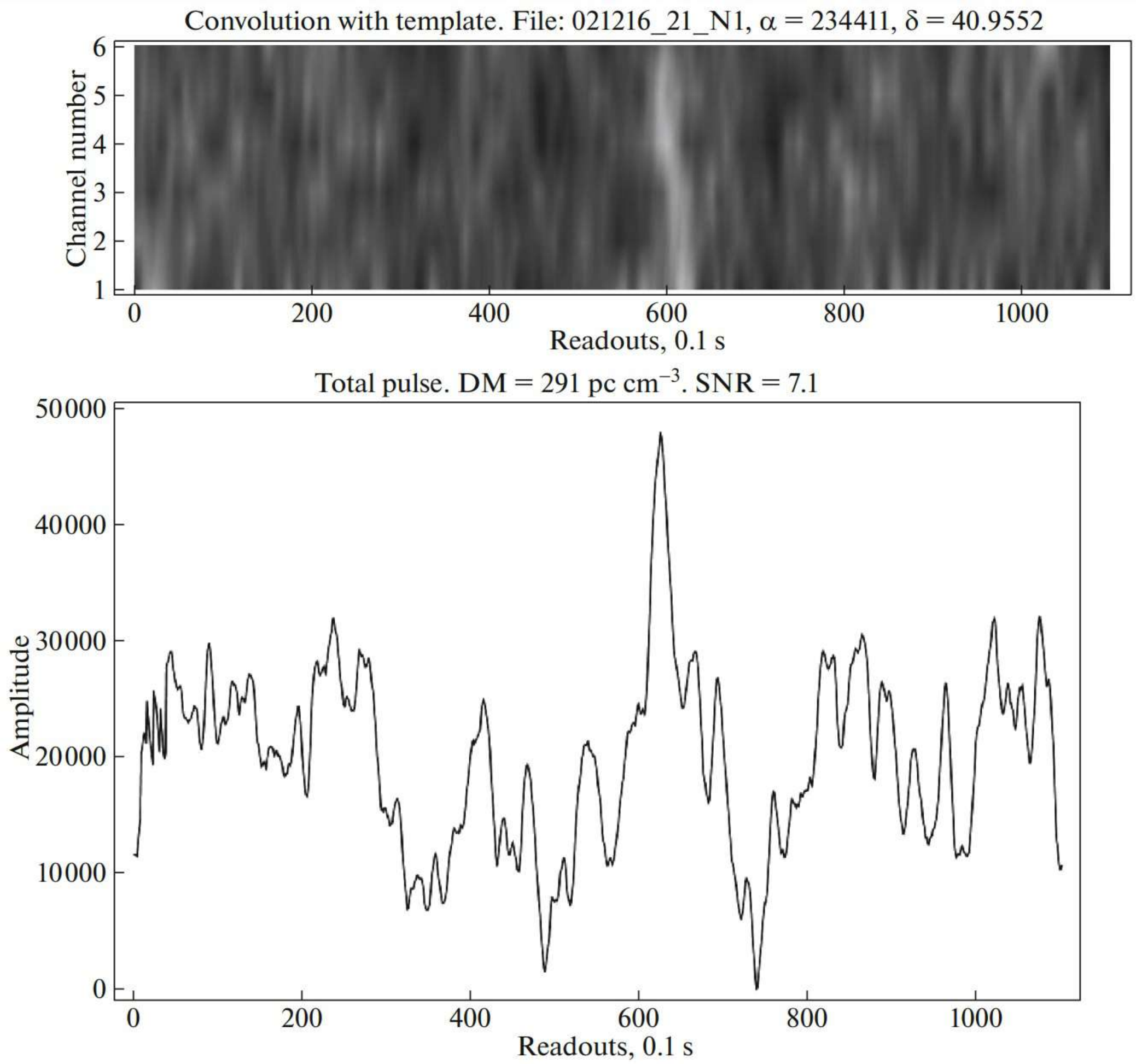}}
	\caption{ Dynamical spectrum of the burst with $DM = 291$ $pc/cm^3$ detected on December 2, 2016. The lower image shows the
	total profile of the pulse. The peak flux density is 0.29 Jy. The Galactic coordinates are $l = 109.48^\circ$, $b =$ -- 20.32$^\circ$.}
	\label{ris:fig1}
\end{figure}

\newpage
\begin{figure}[h!]
	\setcaptionmargin{1.3mm}
	\vbox{\includegraphics[width=0.9\linewidth]{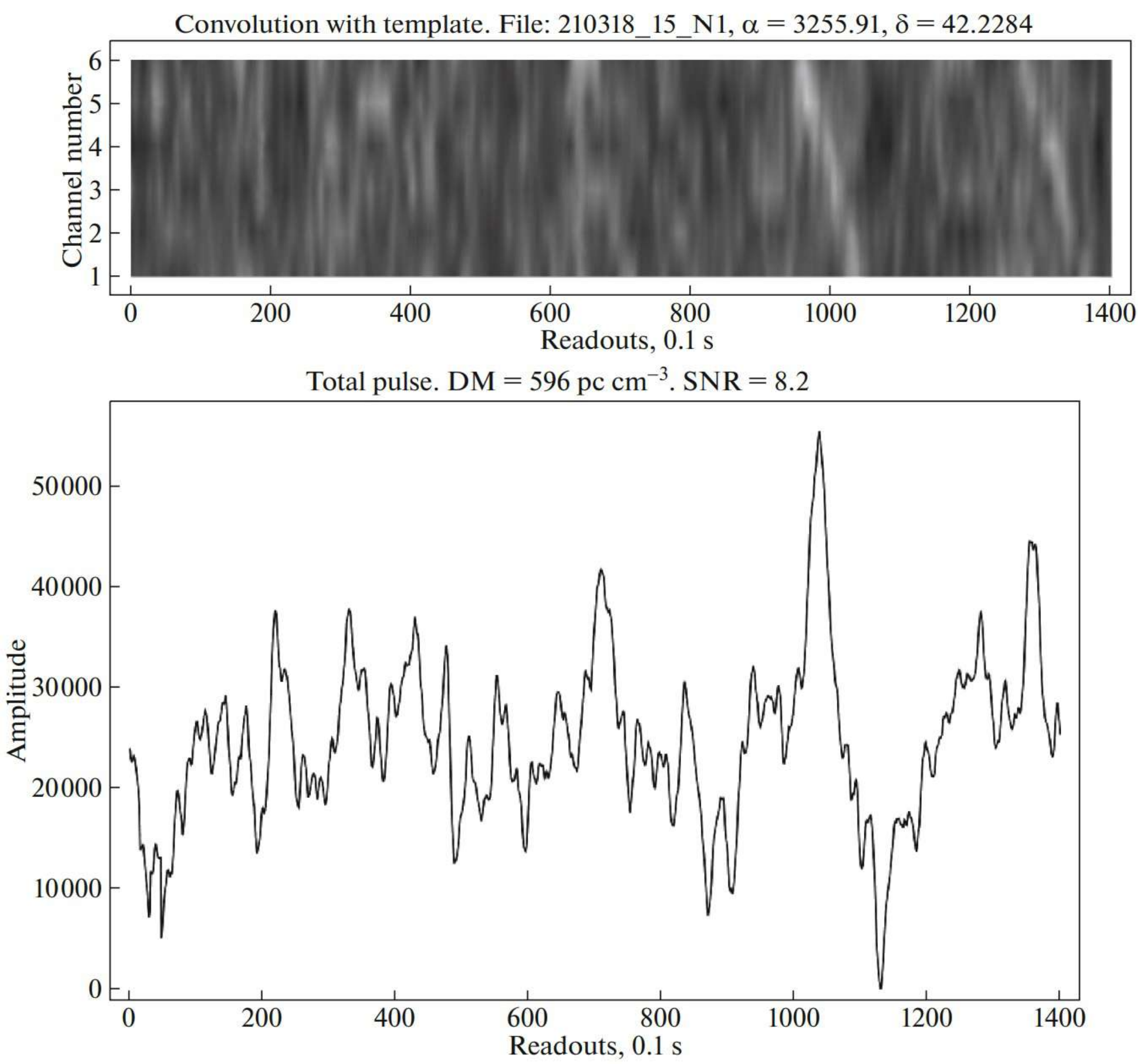}}
	\caption{ Dynamical spectrum of the burst with $DM = 596$ $pc/cm^3$ detected on March 21, 2018. The lower image shows the
	total profile of the pulse. The peak flux density is 0.54 Jy. The Galactic coordinates are $l = 119.46^\circ$, $b =$ -- 20.$72^\circ$.}
	\label{ris:fig1}
\end{figure}

\newpage
\begin{figure}[h!]
	\setcaptionmargin{1.3mm}
	\vbox{\includegraphics[width=1\linewidth]{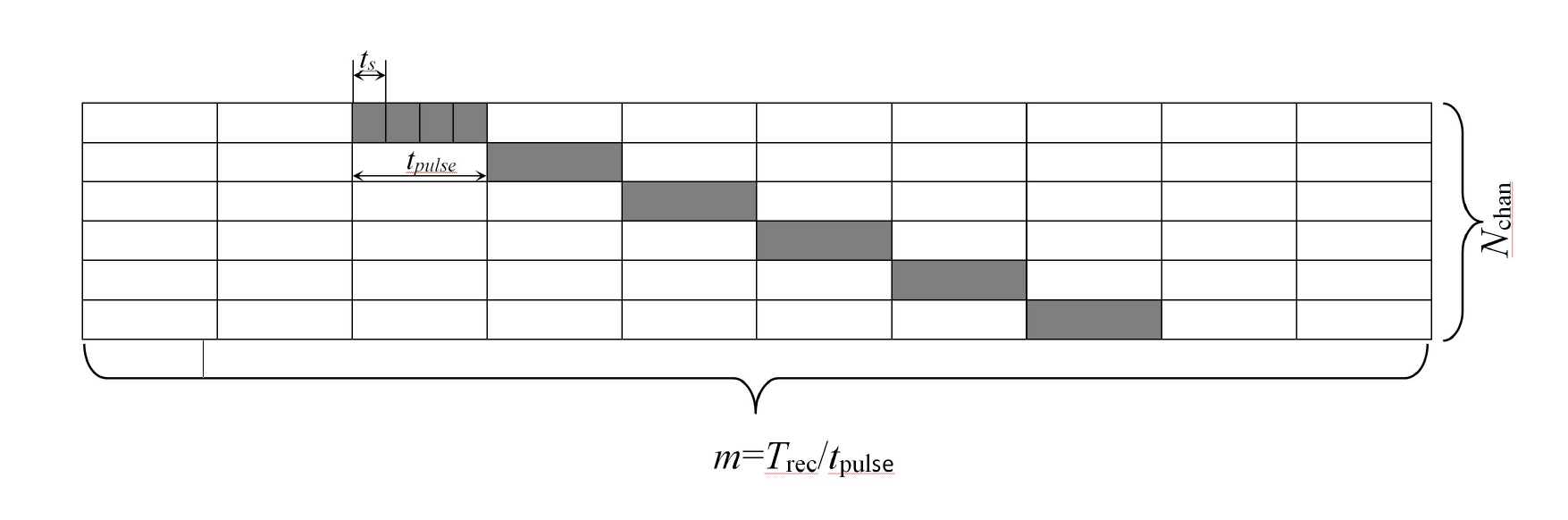}}
	\caption{. Schematic image of the dynamical spectrum for computation of the probability of appearance by chance. $T_{rec}=300\thickspace c$ is the duration of the recording, $N_{chan}$ = 6 the number of frequency channels, $t_s\thickspace=\thickspace 1\thickspace s$ the scattering for the template, and $t_{pulse}$ the characteristic pulse width.}
	\label{ris:fig1}
\end{figure}

\end{document}